\documentstyle[amssymb,aps,12pt]{revtex}

\begin{document}
\title{Spin filtering and magnetoresistance in ballistic tunnel junctions}
\author{J.C. Egues$^*$}
\address{Departamento de F\'{\i}sica e Inform\'{a}tica, Instituto de F\'{\i}%
sica de S\~{a}o Carlos, \ \ \\
Universidade de S\~{a}o Paulo, 13560-970 S\~{a}o Carlos,\\
S\~{a}o Paulo, Brazil.}
\author{C. Gould, G. Richter and L. W. Molenkamp}
\address{Physikalisches Institut, Universit\"{a}t W\"{u}rzburg, Am Hubland,
97074\\
W\"{u}rzburg, Germany}
\date{\today}
\maketitle

\begin{abstract}
We theoretically investigate magnetoresistance (MR) effects in connection
with spin filtering in quantum-coherent transport through tunnel junctions
based on non-magnetic/semimagnetic heterostructures. We find that spin
filtering in conjunction with the suppression/enhancement of the
spin-dependent Fermi seas in semimagnetic contacts gives rise to (i)
spin-split kinks in the MR of single barriers and (ii) a robust beating
pattern in the MR of double barriers with a semimagnetic well. We believe
these are unique signatures for quantum filtering.
\end{abstract}


\newpage

\section{Introduction}

The recent experimental demonstrations of spin-polarized currents in
Mn-based semiconductors \cite{Fiederling}-\cite{jonker} represent a crucial
first step towards understanding spin-dependent transport in these systems
and possibly devising real spintronic devices \cite{Prinz}. So far spin
injection has been verified only at cryogenic temperatures. Low temperature
spin transport is, however, extremelly important as a testing ground for
novel ideas and concepts in the emerging fields of {\em semiconductor}
spintronics and (possibly) quantum computing \cite{burkard}.

The spin-injection experiments involving a semimagnetic layer as a spin
aligner \cite{oestreich} reported to date \cite{Fiederling}-\cite{jonker}
pertain to transport in the {\em diffusive} limit. Moreover, the high
voltages used ($eV\gg $ Fermi energy) make transport highly {\em non-linear}%
. More recently Schmidt {\em et al.}\cite{schmidt} have investigated MR in
diffusive spin transport through a non-magnetic semiconductor (NMS) layer
with dilute-magnetic semiconductor (DMS) contacts. In the regime of linear
response they find a positive MR due to the suppression of one spin channel
in the non-magnetic region.

The ballistic or quantum-coherent limit is another interesting regime in
which to look at spin-polarized transport. In this regime, {\it spin
filtering} \cite{egues} can give rise to a spin-polarized flow. As detailed
in Ref. \cite{egues}, spin filtering in semimagnetic systems is due to
selective electron transmission. The {\em s-d} interaction gives rise to a
spin-dependent potential: spin-up and spin-down electrons see different
barrier heights. Hence one spin-component is blocked while the other is not.
Observe that in diffusive transport {\it spin aligning} \cite{oestreich} due
to spin-flip processes is the dominant mechanism behind the generation of
spin-polarized currents in semimagnetic semiconductors.

Here we theoretically investigate MR for ballistic transport in semimagnetic
heterojunctions\cite{sf} with several arrangements of semimagnetic contacts
and tunnel barriers as shown in Fig. 1(a)-1(h). The idea is to find
signatures of the spin-filtering effect in the MR; so far no experimental
evidence for spin filtering in semimagnetic heterostructures has been
reported. Our current density calculation follows that of Ref.\cite{egues}
properly generalized to account for {\em spin-dependent contacts}. We
calculate the spin-up and spin-down linear \cite{chang} conductances in
terms of the respective transmission coefficients and present explicit
formulas for single- and double-barrier cases.

{\it Findings.} For single-barrier structures with DMS contacts and NMS
barriers, Fig. 1(a),(b), we find that spin filtering and the spin-dependent
changes of the Fermi seas in the contacts give rise to an enhanced and
essentially positive MR, Fig. 2(a). Spin-split kinks in MR are also seen,
Fig. 2(b); these result from subsequent spin-resolved Landau levels crossing
the Fermi surface in the contacts and hence closing the corresponding
conducting channels. For double-barrier systems comprised of DMS contacts
and well with non-magnetic barriers, Fig. 1(c),(d), we find particularly
interesting beating in the MR. This robust feature comes about because of
the {\it significantly enhanced} spin splitting of the resonant levels in
the well; this results from the unique alignement of spin-dependent band
edges in the contacts with the respective bottom of the potential well --
provided by the particular geometry used. For semimagnetic barriers with
non-magnetic contacts, Fig. 1(e),(f), spin filtering suppresses the
Landau-level-induced kinks and makes MR essentially negative, Fig. 3.

The above features markedly contrast with the MR in non-magnetic
heterostructures. We believe the peculiar structures in the MR of
semimagnetic heterojunctions constitute unique signatures of the interplay
between quantum spin filtering and spin-dependent phase-space modulation in
the contacts.

\section{Model system}

Consider a two-terminal geometry with DMS contacts separated by a tunneling
region. In the presence of a magnetic field $B$ along the growth direction $z
$, the transverse motion is quantized into Landau levels \cite{LL}. Along
the field we have parabolic spin-dependent bands $\xi _{\sigma
_{z}}(k_{z},B)=\varepsilon _{\sigma _{z}}(B)+E_{z}(k_{z})$ with $%
E_{z}(k_{z})=\hbar ^{2}k_{z}^{2}/2m_{e}^{\ast }$, $k_{z}$ is the electron
wave vector and $m_{e}^{\ast }$ the effective mass. The spin-dependent band
edges are $\varepsilon _{\sigma _{z}}\equiv \pm \epsilon (B)$, where the
upper sign refers to spin-up electrons and $\epsilon (B)\equiv
x_{eff}|\langle S_{z}\rangle |N_{0}\alpha /2>0$, $x_{eff}=x(1-x)^{12}$ is
the effective Mn concentration (accounting for Mn-Mn antiferromagnetic
pairing), $\langle S_{z}\rangle $ is the 5/2 Brillouin function describing
the thermal average of the Mn spin components, and $N_{0}\alpha $ is the
{\em s-d} exchange constant for conduction electrons.

{\em Current density. }The tunneling region is described by a spin-dependent
transmission coefficient $T_{\sigma _{z}}(E_{z},V,B)$. By extending the
approach in Ref. \cite{egues} to the present case we can write\ the current
density across the tunneling region at zero temperature \cite{temperature}
and for $eV<E_{F}$ as
\begin{eqnarray}
J_{\sigma _{z}}(B) &=&\bar{J}_{0}\hbar \omega _{c}\Bigg\{\int%
\limits_{0}^{E_{F}^{\sigma _{z}}-\frac{1}{2}\hbar \omega _{c}}\left[
\mathop{\rm int}%
\left( \frac{E_{F}^{\sigma _{z}}-E_{z}}{\hbar \omega _{c}}-\frac{1}{2}%
\right) +1\right] T_{\sigma _{z}}(E_{z},V,B)dE_{z}-  \nonumber \\
&&\int\limits_{0}^{E_{F}^{\sigma _{z}}-\frac{1}{2}\hbar \omega _{c}-eV}\left[
\mathop{\rm int}%
\left( \frac{E_{F}^{\sigma _{z}}-eV-E_{z}}{\hbar \omega _{c}}-\frac{1}{2}%
\right) +1\right] T_{\sigma _{z}}(E_{z},V,B)dE_{z}\Bigg\},  \label{eq1}
\end{eqnarray}%
where $\bar{J}_{0}\equiv em_{e}^{\ast }/4\pi ^{2}\hbar ^{3}$, $%
E_{F}^{\uparrow ,\downarrow }=E_{F}\mp \epsilon (B)$, $\omega _{c}$ is the
cyclotron frequency, and $%
\mathop{\rm int}%
(x)$ denotes the largest integer smaller than or equal to $x.$ Equation (\ref%
{eq1}) is the $B\neq 0$ Tsu-Esaki formula with spin-dependent trasmission
coefficientes.

\subsection{Linear response}

{\em Linear spin-dependent conductance.} To determine the linear conductance
we linearize Eq. (\ref{eq1}). Taylor expanding (\ref{eq1}) around $eV=0$ and
using
\begin{equation}
\frac{d%
\mathop{\rm int}%
(x)}{deV}=\frac{d%
\mathop{\rm int}%
(x)}{dx}\frac{\partial x}{\partial eV}=-\frac{1}{\hbar \omega _{c}}%
\sum\limits_{n}\delta (x-n),\text{ }  \label{eq3}
\end{equation}%
where {\em n} is an integer and $x=\frac{E_{F}^{\sigma _{z}}-eV-E_{z}}{\hbar
\omega _{c}}-\frac{1}{2}$, we find to linear order in $eV$
\begin{equation}
J_{\sigma _{z}}(B,E_{F},eV)=\bar{J}_{0}\hbar \omega
_{c}\sum\limits_{n=0}^{n_{0}^{\sigma _{z}}}T_{\sigma _{z}}\left(
E_{F,n}^{\sigma _{z}},0,B\right) eV,  \label{eq5}
\end{equation}%
with $n_{0}^{\sigma _{z}}=%
\mathop{\rm int}%
\left( \frac{E_{F}^{\sigma _{z}}}{\hbar \omega _{c}}-\frac{1}{2}\right) $
and $E_{F,n}^{\sigma _{z}}\equiv E_{F}^{\sigma _{z}}-\left( n+\frac{1}{2}%
\right) \hbar \omega _{c}$. The spin-dependent linear conductance per unit
area $G_{\sigma _{z}}(B,E_{F})/A\equiv J_{\sigma _{z}}(B,E_{F},eV)/V$ is
\begin{equation}
G_{\sigma _{z}}(B,E_{F})/A=\frac{e^{2}m_{e}^{\ast }}{4\pi ^{2}\hbar ^{3}}%
\hbar \omega _{c}\sum\limits_{n=0}^{n_{0}^{\sigma _{z}}}T_{\sigma
_{z}}\left( E_{F,n}^{\sigma _{z}},0,B\right) .  \label{eq6}
\end{equation}%
Observe that the transmission coefficient in Eq. (\ref{eq6}) is calculated
{\em in the absence} of any external potential; this is just the general
philosophy of linear response: the response depends upon only the system
configuration in equilibrium. It is straightforward to verify that Eq. (\ref%
{eq6}) reduces to the well known result\cite{duke}
\begin{equation}
G_{0}(E_{F})/A=\frac{e^{2}m_{e}^{\ast }}{4\pi ^{2}\hbar ^{3}}%
\int\limits_{0}^{E_{F}}dE_{z}T_{0}\left( E_{z}\right) ,  \label{eq7}
\end{equation}%
in the $B=0$ limit, where $G_{0}(E_{F})\equiv G_{\uparrow
}(E_{F},0)=G_{\downarrow }(E_{F},0)$ and $T_{0}\left( E_{z}\right) \equiv
T_{\uparrow }\left( E_{z},0\right) =T_{\downarrow }\left( E_{z},0\right) $.

\subsection{Magnetoresistance}

\noindent Let $R$ ($R_{0}$) and $G$ ($\tilde{G}_{0}=2G_{0}$) be the {\em %
total } resistance and conductance, respectively, in the presence (absence)
of magnetic fields. The MR of the system is defined by 
$\frac{\Delta R}{R_{0}}=\frac{R}{R_{0}}-1=\frac{\tilde{G}_{0}}{G}-1=\frac{%
2G_{0}(E_{F})}{G_{\uparrow }(B,E_{F})+G_{\downarrow }(B,E_{F})}-1$.
Hence
\begin{equation}
\frac{\Delta R}{R_{0}}=\frac{2\int\limits_{0}^{E_{F}}dE_{z}T_{0}(E_{z})}{%
\hbar \omega _{c}\left[ \sum\limits_{n=0}^{n_{0}^{\uparrow }}T_{\uparrow
}\left( E_{F,n}^{\uparrow },B\right) +\sum\limits_{n=0}^{n_{0}^{\downarrow
}}T_{\downarrow }\left( E_{F,n}^{\downarrow },B\right) \right] }-1.
\label{eq10}
\end{equation}%
In deriving Eq. (\ref{eq10}) we assume a negligible contact resistance.

\subsection{Transmission coefficients}

\noindent {\em Single- and double-barrier potentials.} In order to determine
$\Delta R/R_{0}$ from Eq. (\ref{eq10}) we only need the transmission
coefficients for {\em zero} applied voltage (equilibrium configuration)
since we are in the linear response regime. For a single barrier of width $%
L_{b}$ and height $V_{b}$ we readily find \cite{egues}
\begin{equation}
T_{\uparrow ,\downarrow }^{SB}(E_{F,n}^{\uparrow ,\downarrow },B)=\left\{ 1+%
\frac{\sinh ^{2}\left[ \sqrt{\frac{2m_{e}^{\ast }(V_{\uparrow ,\downarrow
}-E_{F,n}^{\uparrow ,\downarrow })}{\hslash ^{2}}}L_{b}\right] }{4\left(
\frac{E_{F,n}^{\uparrow ,\downarrow }}{V_{\uparrow ,\downarrow }}\right)
\left( 1-\frac{E_{F,n}^{\uparrow ,\downarrow }}{V_{\uparrow ,\downarrow }}%
\right) }\right\} ^{-1},  \label{eq11}
\end{equation}
with $V_{\uparrow ,\downarrow }=V_{b}\mp \epsilon (B)$ and $%
E_{F,n}^{\uparrow ,\downarrow }=E_{F}\mp \epsilon (B)-(n+1/2)\hbar \omega
_{c}$.

For a symmetric DMS-contacted double-barrier structure with a semimagnetic
well of width $L_{w}$ and NMS barriers of width $L_{b}$ and height $V_{b}$,
we have
\begin{eqnarray}
T_{\uparrow ,\downarrow }^{DB}(E_{F,n}^{\uparrow ,\downarrow },B) &=&\Bigg|%
\left[ \cosh \left( \kappa _{n}^{\uparrow ,\downarrow }L_{b}\right) -i\frac{%
2E_{F,n}^{\uparrow ,\downarrow }-V_{\uparrow ,\downarrow }}{2\sqrt{%
E_{F,n}^{\uparrow ,\downarrow }(V_{\uparrow ,\downarrow }-E_{F,n}^{\uparrow
,\downarrow })}}\sinh \left( \kappa _{n}^{\uparrow ,\downarrow }L_{b}\right) %
\right] ^{2}e^{-ik_{n}^{\uparrow ,\downarrow }L_{w}}+  \nonumber \\
&&\frac{V_{\uparrow ,\downarrow }^{2}}{4E_{F,n}^{\uparrow ,\downarrow
}(V_{\uparrow ,\downarrow }-E_{F,n}^{\uparrow ,\downarrow })}\sinh
^{2}\left( \kappa _{n}^{\uparrow ,\downarrow }L_{b}\right)
e^{ik_{n}^{\uparrow ,\downarrow }L_{w}}\Bigg|^{-2},  \label{eq12}
\end{eqnarray}%
where $k_{n}^{\uparrow ,\downarrow }=\sqrt{2m_{e}^{\ast }E_{F,n}^{\uparrow
,\downarrow }}/\hbar $ and $\kappa _{n}^{\uparrow ,\downarrow }=\sqrt{%
2m_{e}^{\ast }(V_{\uparrow ,\downarrow }-E_{F,n}^{\uparrow ,\downarrow })}%
/\hbar $. Equations (\ref{eq11}) and (\ref{eq12}) hold for $V_{\uparrow
,\downarrow }\geqslant E_{F,n}^{\uparrow ,\downarrow }$. In what follows we
discuss some plots of $\Delta R/R_{0}$ vs $B$ for both single- and
double-barrier heterostructures. In the subsequent graphs we use $x=0.06$
and $m_{e}^{\ast }=0.16m_{0}$. Fermi energies and potential heights (and
widths) are shown in the figures.

\section{Results}

{\em Results.} Figure 2 shows the MR of a DMS/NMS/DMS structure, Fig.
1(a),(b), for several NMS barrier widths. Observe that the MR is {\it %
enhanced and mostly positive} for wider barriers and high fields, as
compared to the non-magnetic case (dashed lines). These features are due to
spin filtering resulting from the relative change of the band edges in the
DMS contacts together with the concomitant reduction and increase of the
Fermi seas for spin-up and spin-down electrons, respectively. Positive MR is
expected for high enough fields above which the spin-up channels are
unavailable. Here, for $B>~4.7$ T the spin-up conductance is identically
zero; see the kink and the steep rise of $\Delta R/R_{0}$ around $B=4.7$ T
for all barrier widths. Note that because the conductance of the system is
essentially a sum over spin-resolved Landau channels smaller than $%
n_{0}^{\uparrow ,\downarrow }$, Eq. (\ref{eq6}), the abrupt closing of
channels manifests itself directly in $\Delta R/R_{0}$ via Eq. (\ref{eq10}).
Figure 2(b) illustrates the connection between the shut down of
spin-resolved transmission channels and the kinks in the MR more clearly.
Note that the more abruptly a channel shuts down, the steeper $\Delta
R/R_{0} $ rises in its vicinity. Note also that the spin-resolved Landau
levels give rise to spin-split kinks in the MR. The inset shows that the
interplay between spin-dependent phase space in the DMS contacts and spin
filtering can also lead to negative MR for larger barriers and smaller
fields. All the above features contrast with the no s-d exchange case
(dashed lines).

Figure 3 shows a plot similar to that in Fig. 2 but for a single DMS barrier
with NMS contacts\cite{ref14}. Note that the MR is now {\it mostly negative}%
. Here this happens entirely because of quantum-coherent spin filtering \cite%
{egues}. As the magnetic field increases the spin-up (spin-down) electrons
see a larger (smaller) barrier. The corresponding exponential suppression of
$T_{\uparrow }$ and the concomitant enhancement of $T_{\downarrow }$ is
asymmetric. Because the wave function penetration in the DMS barrier is
larger for spin down than for spin up electrons, the former see a stronger
{\it s-d} modulation of the barrier height than the latter; $T_{\downarrow }$
then increases faster than $T_{\uparrow }$ decreases. Hence the total
trasmission coefficient $T_{\uparrow }+T_{\downarrow }$ increases as
compared to the $B=0$ case thus leading to $\Delta R/R_{0}<0$. This effect
is more pronounced for larger barrier widths. For narrower barriers $\Delta
R/R_{0}$ vs $B$ also presents kinks and regions of positive and negative MR
(see curves for the 5 and 10 nm barriers).

Figure 4 shows our results for a symmetric DMS-contacted double-barrier
system with a semimagnetic well and non-magnetic barriers. The $B=0$
configuration is chosen so that there is a resonant level below $E_{F}$ for
the parameters used. A remarkable feature in Fig. 4 is the beating in the MR %
\cite{beating}. As shown in the lower part of this figure, this beating is
directly related to the peculiar overlap of the s-d spin-split transmission
channels. The unique pattern in $\Delta R/R_{0}$ is made particularly
noticeable by the geometry used -- which effectively enhances {\it s-d}
induced features. By considering semimagnetic contacts {\it and} well, we
are essentially forcing the spin-resolved resonant states in the well to
have larger spin splittings, Fig. 1(c),(d). This happens because the
spin-dependent bottom of the potential well lines up with the corresponding
spin-split band edges in the DMS contacts. In simpler terms, we are
referring the spin-resolved states to spin-split origins in the contacts.
The end result is indeed a larger effective spin-splitting of the resonant
levels. Note that the DMS contacts do not need to be fully spin polarized.

{\em Feasibility.} Recent advances in DMS materials technology make the
short term realization of the structures suggested in this paper realistic %
\cite{e-e}. For instance, combined DMS-NMS heterostructures have already
been demonstrated \cite{schmidt} in ZnBeMnSe/ZnBeSe quartenary materials.
The band offsets and doping densities achievable by varying the
concentrations of these compounds are flexible enough to produce the
potential profiles and Fermi energies considered here. Furthermore, we
estimate the electron coherence lenghts in these materials to be longer than
the total length of our structures thus allowing for quantum-coherent
transport.

{\it Summary.} We have shown that DMS-contacted tunnel junctions display
very peculiar MR due to quantum-coherent spin filtering in conjunction with
the reduction/enhancement of the spin-dependent Fermi seas in the contacts.
The features in the MR reported here (e.g., beating and spin-split kinks)
are unique signatures of spin filtering in ballistic transport. We expect
these effects to be easily resolved experimentally.

This work was supported by FAPESP and Deutsche Forschungsgemeinschaft (DFG,
SFB 410). J.C.E. acknowledges the kind hospitality at the Physikalisches
Institut, Universit\"{a}t W\"{u}rzburg, where part of this work was
developed, and financial support for his visit.

\bigskip \noindent$^*$Electronic address: egues@if.sc.usp.br

\newpage

\begin{figure}[t]
\caption{ Spin-dependent potential profiles for several tunneling
structures. In DMS-contacted geometries with applied {\it B}, the
{\it s-d} exchange interaction shifts the spin-up and spin-down
band edges upwards and downwards, respectively, relative to the
$B=0$ case, thus creating partially spin-polarized reservoirs
[only subbands for the lowest Landau levels are shown: $\hbar
\protect\omega _{c}/2 \pm \protect\epsilon (B)+\hbar
^{2}k_{z}^{2}/2m_e^{\ast }$]. This is illustrated in (a) and (b)
for a single barrier and in (c) and (d) for a double barrier. A
single DMS barrier with NMS contacts is shown in (e) and (f). Note
that now the {\it s-d} interaction modulates only the barrier
height in a spin-dependent fashion; the electron reservoirs are
here unpolarized. The NMS-contacted double-barrier structure with
a DMS well is shown in (g) and (h). The left panel shows only the
lowest Landau level bands (contacts) for $B\neq 0$. A particularly
large spin splitting is attained by using a DMS-contacted
double-barrier geometry with semimagnetic well in contrast to the
non-magnetic contact case: $\Delta ^{\prime }-\protect\delta \approx \Delta=2%
\protect\epsilon (B) $. }
\label{fig1}
\end{figure}

\begin{figure}[tbp]
\caption{Magnetic-field dependence of $\Delta R/R_0$ (a) and transmission
coefficients (b) for single NMS barriers with DMS contacts, Fig. 1(a).
Exchange-induced filtering in conjunction with the corresponding reduction
of the spin-up phase space below $E_F$ in the DMS contacts, gives rise to an
enhanced positive MR for larger barrier widths as compared to the no-s-d
case (dashed lines). The MR is slightly negative for small fields and larger
barriers (inset). The usual kinks due to Landau-level quantization are also
present; however, they are now more pronounced and spin resolved, (b). We
can also see the kinks are due to the subsequent shut down of spin-resolved
transmission channels as the field is increased. Note that current should be
fully spin polarized for $B > 4.7$ T. }
\label{fig2}
\end{figure}

\begin{figure}[tbp]
\caption{Same as Fig. 2(a) but for a NMS/DMS/NMS structure [see Fig. 1(e)].
Much in contrast to the DMS-contacted barrier case in Fig. 2(a) and the no
s-d exchange case (dashed lines), spin filtering here gives rise to enhanced
negative $\Delta R/R_0$ for wider barriers. Narrower DMS barriers show
smaller MR and kinks similar to the no-s-d case; however, the Landau levels
here are not spin resolved.}
\label{fig3}
\end{figure}

\begin{figure}[tbp]
\caption{Beating in the MR of a double-barrier system with both DMS well and
contacts [see Fig. 1(c)]. The beating feature in the MR response is due to
the overlap of many {\it s-d} spin-resolved transmission channels, as
clearly shown in the lower part of the figure. The large spin splitting of
the spin-resolved resonant level in the well -- unique to a geometry in
which, for each spin component, the corresponding bottom of the potential in
the well is aligned with the respective conduction band edge in the contacts
-- makes this beating very pronounced; a strong signature of
quantum-coherent spin-resolved transport in double barriers.}
\label{fig4}
\end{figure}


\begin{references}
\bibitem{temperature} As usual, we consider the electrons at $T=0$ (sharp
Fermi function) and the paramagnetic impurities at $T=4.2$ K.
\bibitem{ref14} The formulas for the conductances, magnetoresistances, and
transmission coefficients for NMS-contacted structures are similar to those
for the DMS-contacted ones. But now $n_{0}^{\uparrow }=n_{0}^{\downarrow
}=n_{0}=%
\mathop{\rm int}%
(E_{F}/\hbar \omega _{c}-1/2),$ $E_{F,n}^{\uparrow ,\downarrow }\equiv
E_{F}-(n+1/2)\hbar \omega _{c}$, and $V_{\uparrow ,\downarrow }=V_{b}\pm
\epsilon (B)$ (i.e., ``$\mp $'' in the former definition of $V_{\uparrow
,\downarrow }$ becomes ``$\pm $'' here).
\end{references}
\end{document}